\documentclass[12pt]{iopart}

\usepackage{graphicx}
\usepackage{color}

\begin{document}

\title[General inner products]{General inner products for energy eigenstates}

\author{J. Julve}

\address{IMAFF, Consejo Superior de Investigaciones
Cient\'\i ficas, Serrano 113 bis, Madrid 28006, Spain}

\ead{julve@iff.csic.es}

\author{S. Turrini}

\address{Dipartimento di Fisica,
Universit\`a di Bologna, Italy}

\ead{turrini@bo.infn.it}

\author{F. J. de Urr\'\i es}

\address{Universidad de Alcal\'a de Henares, Spain}

\ead{fernando.urries@uah.es}

\begin{abstract}
The features of the inner products between all the types of real
and complex-energy solutions of the $\rm Schr\ddot{o}dinger$
equation for 1-dimensional cut-off quantum potentials are worked out
using a Gaussian regularization. A general Master Solution is
introduced which describes any of the above solutions as particular
cases. From it, a Master Inner Product is obtained which yields all
the particular products. We show that the Outgoing and the
Incoming Boundary Conditions fully determine the location of the
momenta respectively in the lower and upper half complex plane even
for purely imaginary momenta (anti-bound and bound solutions).


\submitto{Journal of Physics A: Mathematical and General}

\end{abstract}

\maketitle

\section{Introduction}

The (time-independent) $\rm Schr\ddot{o}dinger$ Equation (SE)
has a variety of solutions according to the Boundary Conditions
we impose on them. Besides the physical bound
eigenstates ($\tilde\phi_i$ with discrete negative real
energy spectrum and positive purely imaginary momenta) we have
marginally physical ones (continuum of positive real energy
scattering solutions $\psi_E$), complex-energy
solutions (resonances $u_n$), the also unphysical
anti-bound states ($\phi_i$ with negative purely imaginary
momenta), and, finally, a complex-energy continuum of "background" states
($u_{\cal E}$). Out of these states, only the bound states have a finite
norm (that is, $\tilde\phi_i\in{\cal L}^2$, the Hilbert space).

The properties and uses of the real-energy bound and scattering (Dirac)
states are a standard textbook topic on foundations of quantum mechanics.
Also the resonances (known as Gamow states \cite{Gamow}) have deserved extensive attention
in the literature as descriptive of decaying states. However they feature a spatially
divergent asymptotic behaviour, which leads to infinite norms and seemingly
divergent inner products. This problem has historically hampered the study of
these solutions and spurred several attempts to circumvent it by trying different
regularization prescriptions. Examples of this are analytical continuations
of the inner products in the complex momentum plane, the External Complex Scaling of
the space coordinate, and the introduction of convergence factors in the space integrals.
The interested reader may find some bibliographical cues on these proposals in
\cite{Nicolaides} \cite{Gadella} \cite{Julve1}. In \cite{Julve1} the Gaussian
regularization has been used to achieve new results on the computation of the inner
products involving resonances.

In this paper we complete our previous results, obtained for a general
1-dimensional potential with finite support, by considering the full set of solutions of
the SE including the anti-bound and the "background" states.

In Section 2, we first characterize the solutions according to the Boundary
Conditions (BCs) they obey, having the Outgoing (OBCs) and Incoming (IBCs)
Boundary Conditions a chief role. We will show that also the bound and anti-bound
solutions fit in the ensuing classification. We then introduce a general form
of the solution to the SE as a Master Solution (MS) from which all the
solutions above stem as particular cases.

In Section 3 we exploit this Master Solution to obtain a compact Master
Inner Product (MIP) from which the table of the particular inner products
between all the solutions can be worked out. This general framework lets us to
see how the different kinds of orthogonality, namely Kronecker-$\delta$, Dirac-$\delta$,
or other, stem from a common expression.

In Section 4 the Conclusions are drawn. The details of some calculations and the
basic integral formulas are deferred to Appendices.

\section{Solutions and Boundary Conditions}

We consider the 1-D time-independent $\rm Schr\ddot{o}dinger$
equation
\begin{equation}
[\frac{\partial^2}{\partial x^2}+p^2-2mV(x)]\;\psi(x,p) =0
\end{equation}
where we have used units such that $\hbar=1$, and a cut-off potential
$V(x)$ describing a general barrier with support in the compact
interval [0,\emph{L}]. We also assume that the potential features
local negative minima (wells) such that bound states occur.

Besides the better known scattering \emph{in} and \emph{out} solutions
for continuous real energy $E=p^2/2m\,>0$ and the bound and anti-bound
states (discrete real $E_i<0)$, one has resonant solutions satisfying
the Siegert Boundary Conditions \cite{Siegert}, and the more general continuous
complex-energy "background" solutions. Here we briefly highlight some of
their features and introduce our systematics together with some notation.
We also stress that outside [0,\emph{L}] the solutions already adopt their
Asymptotic Form (AF) for $x\rightarrow\pm\infty$, the use of which may be a
convenient alternative to their BCs at $x=0,L$ (or at $x=\pm\infty$).
The BCs maybe algebraic or differential
and the general setup is that imposing one BC leaves the functional degree
of freedom corresponding to a continuum (energy or momentum) spectrum, as
is the case of the scattering and background solutions, whereas two BCs
over-determine the problem yielding a discrete spectrum, whether of real
energies (bound and anti-bound solutions) or complex energies (resonances).
Graphical representations of the location of the momenta of the different
solutions in the complex-momentum plane can be found in \cite{Julve1}.

\subsection{Bound and anti-bound states}

The bound states $\tilde\phi_i$ are the only solutions $\in{\cal L}^2$ and
they are characterized by the AF
\begin{equation}
\tilde\phi_i(x)=\left\{
\begin{array}{ll}
R_i\,e^{\tilde q_ix}&\;,\quad x\leq 0\\
T_i\,e^{-\tilde q_ix}&\;,\quad x\geq L
\end{array}\right.
\end{equation}
where $\tilde q_i>0$. This is equivalent to imposing the BCs
\begin{equation}
\partial_x\tilde\phi|_{x=0}=\tilde q\,\tilde\phi(0),\quad \quad \partial_x\tilde\phi|_{x=L}=-\tilde q\,\tilde\phi(L)\quad(\tilde q>0)\,,
\end{equation}
which imply the weaker BCs $\tilde\phi_i(\pm\infty)=0$.
The anti-bound solutions instead correspond to the mirror values $q_i=-\tilde q_i<0$
in (2) and (3), then becoming  $\notin{\cal L}^2$ (thus dubbed unphysical), and have the
same negative energies $E_i=-q_i^2/2m$.

The AF of both types of solutions maybe viewed as plane waves (say $\rme^{\rmi p_ix}$ in
the right side of the barrier) with purely imaginary "momenta" $p_i$ which may lie in
the positive imaginary axis ${\cal I}_+$ (bound states) or in the negative one ${\cal I}_-$
(anti-bound solutions). This interpretation makes these states akin to the resonances.
The \emph{tilde} notation we have adopted for the bound solutions reflects their relationship
with the \emph{Incoming} resonances below.

\subsection{Resonances}

They are characterized by the
Siegert homogeneous \emph{Outgoing} BCs (OBCs)
\begin{equation}
\partial_x u\!\!\mid_{x=0}\,=-\rmi p\,u(0) \;\;,\;\;
\partial_x u\!\!\mid_{x=L}\,=\rmi p\,u(L)\;,
\end{equation}
from where (3) is recovered as the particular case of purely imaginary momenta $p=\rmi\tilde q$.
Solutions $u_n(x)$, called proper resonances,
exist for a denumerable set of isolated complex values $p_n$ of $p\;$ (with corresponding
complex energies $Z_n=p^2_n/2m$) lying inside the octants close to the real axis
(i.e. $|{\rm Re}\,p_n|>|{\rm Im}\,p_n|$) in the lower half complex plane ${\cal C}_-$
(i.e. ${\rm Im}\,p_n<0$),
and occupy symmetrical positions with respect to the imaginary axis.
It is customary to label them as $p_n$ ($n=1,2,...$) when ${\rm Re}\,p_n
>0$ and as $p_{-n}\equiv-p_n^*$ their symmetric ones, some times called anti-resonances.
The real parts ${\rm Re}\,p_n$ tend to be spaced regularly for
increasing $|n|$ , while $|\,{\rm Im}\,p_n|$ grows slowly
\cite{Nussenzveig}.

BCs with reversed sign of $p$ in (4) (call them IBCs) correspond
to \emph{Incoming} solutions $\tilde{u}_ n(x)$. For real potentials $V(x)$ one
has that $\tilde{u}_ n(x)=u^*_n(x)$ , and the momenta
$\tilde{p}_n=p^*_n$ lie in the upper half complex plane ${\cal C}_+$. We denote
with $|\tilde{z}_n\rangle$ these states. As stressed above, the bound states
maybe viewed as Incoming resonances with zero real part momenta, whereas
the anti-bound solutions would be particular cases of the Outgoing resonances.

The OBCs are equivalent to imposing the asymptotic form
\begin{equation}u_n(x)=\left\{
\begin{array}{ll}
R_n\,e^{-{\rm
i}p_nx}&\quad,\quad x\leq 0\\
T_n\,e^{{\rm i}p_nx}&\quad,\quad x\geq L
\end{array}\right.
\end{equation}
where the amplitudes $R_n$ and $T_n$ differ by a phase and are
defined up to a global arbitrary normalization factor. An immediate
consequence is that the norm is even more divergent than for the Dirac
states, causing both kinds of states not to belong to ${\cal L}^2$.
Moreover, inner products involving resonant (as well as any other
non-normalizable) states, maybe expected to be generally divergent
as well, so that their actual calculation requires a regularization.

\subsection{Scattering solutions}

The continuum spectrum of the scattering states maybe characterized by
just one BC, which can be taken at $x=0$ or $x=L$ and chosen to be of
the Outgoing or of the Incoming type, with an obvious corresponding AF.
Let us call SBCs any of these choices.
For instance, a right-moving $in$ state obeys only the second OBC
in (4) with $p>0$. This is equivalent to imposing the following form
\begin{equation}
\psi^+_r(x)= N(p)\left\{
\begin{array}{ll}
\rme^{{\rm i}px}+R(p)\,\rme^{-{\rmi}px}&\quad,\quad x\leq 0\\
N^{-1}\,\psi^+_r(x)&\quad,\quad 0<x<L\\
T(p)\,\rme^{{\rm i}px}&\quad,\quad x\geq L
\end{array}\right.
\end{equation}
where the form of the solution in $[0,L]$ depends on $V(x)$, and we adopt
the $\delta$-energy wave function normalization $N(p)=\sqrt{m}/\sqrt{2\pi p}$ .
We stress here that the customary $in$-$out$ notation for
the scattering states is misleading as referring to the BCs they obey,
as long as OBCs correspond to $in$ states, like (6), whereas IBCs
yield the \emph{out} states $\psi^-_{r,\,l}(x)$.

\subsection{Complex extension, background states, and Master Solution}

The continuation to the complex plane $p\,$ of the scattering solutions
$\psi^{\pm}_{r,\,l}(x)$ is the guiding thread of this paper. For a general
complex momentum $p=z$ one directly has the so-called $background$ solutions.
For instance, out of (6) one has
\begin{equation}
\Psi_{\cal E}^{+r}(x)=
{\cal N}(z)\left\{
\begin{array}{ll}
\rme^{{\rm i}zx}+R(z)\,\rme^{-{\rmi}zx}&\quad,\quad x\leq 0\\
{\cal N}^{-1}\,\Psi_{\cal E}^{+r}(x)&\quad,\quad 0<x<L\\
T(z)\,\rme^{{\rm i}zx}&\quad,\quad x\geq L
\end{array}\right.
\end{equation}
where ${\cal N}(z)=\sqrt{m}/\sqrt{2\pi z}$ and ${\cal E}=\frac{z^2}{2m}$ is the complex energy.
We notice that, for $z$ almost everywhere in ${\cal C}$, the absolute value of (7) diverges for
$x\rightarrow+\infty$, $x\rightarrow-\infty$, or both, because of the presence of the
asymptotic reflected wave. As an example, we see that $\Psi_{\cal E}^{+r}(z;x)$ diverges
for $x\rightarrow\pm\infty$ when $z\in{\cal C}_-$, and only for $x\rightarrow-\infty$
when $z\in{\cal C}_+$ .

The solution (7) features reflection, transmission (and others in the region $[0,L]$)
amplitudes, which share a common denominator ${\cal D}_+(z)$, which is related to the (radial)
Jost functions in the case of the (s-wave) 3-D scattering. It goes similarly for
the general solution  $\Psi^{\pm r,\,l}_{\cal E}(x)$. Generally, ${\cal D}(z)$ has
complex zeroes $z=p_n$ (resonances) and pure imaginary ones $z=p_i$ (bound and
anti-bound momenta). We then write $R={\cal R}(z)/{\cal D}_+(z)$ , $T={\cal T}(z)/{\cal D}_+(z)$
, etc., where the numerators are holomorphic functions of $z$. Multiplying (7) by ${\cal D}_+(z)$
we then define the Master Solution
\begin{equation}
\Phi_{\cal E}^{+r}(x)=
{\cal N}(z)\left\{
\begin{array}{ll}
{\cal D}_+(z)\,\rme^{{\rm i}zx}+{\cal R}(z)\,\rme^{-{\rmi}zx}&\quad,\quad x\leq 0\\
{\cal N}^{-1}\,\Phi_{\cal E}^{+r}(x)&\quad,\quad 0<x<L\\
{\cal T}(z)\,\rme^{{\rm i}zx}&\quad,\quad x\geq L
\end{array}\right.
\end{equation}
For real $z=p>0$ , ${\cal D}_+(p)$ is regular and one one has $\Phi_{E}^{+r}(x)={\cal
D}_+(p)\,\psi^+_r(x)$ . Instead, (8) readily yields
the resonant and (we shall see) anti-bound solutions, respectively at $z=p_n$ and $z=p_i$ .

As an example, the denominator ${\cal D}_+(p)$, specific to the \emph {in} states
$r,l$ , and its partner ${\cal D}_-(p)$ of the \emph{out} states, have
the following form in the case of a square potential well with depth $V<0$ and width $L$ :
\begin{equation}
\begin{array}{ll}
{\cal D}_+(p)&=(p+\hat p)^2\rme^{-{\rm i}\hat p L}-(p-\hat p)^2\rme^{{\rm i}\hat p L}\\
{\cal D}_-(p)&=(p+\hat p)^2\;\rme^{{\rm i}\hat p L}-(p-\hat p)^2\;\rme^{-{\rm i}\hat
p L}
\end{array}
\end{equation}
where $\hat p =\sqrt{p^2-2mV}$. For real $p$ they are related
by complex conjugation, which is equivalent to changing
the sign of the momenta $p\rightarrow -p$ , $\hat p\rightarrow-\hat
p$ . Notice that the derivation of (9) from SBCs is valid only for $\hat p\neq0$,
so that the value $\hat p=0$ (to which it corresponds $p=p_\pm\equiv\pm\,\rmi\sqrt{2m|V|}$ and $E=V$)
is a meaningless zero of (9). In fact, $p_\pm$ are the only points of ${\cal C}$ for which no eigenfunction of (1) obeying SBCs exists, whereas $E=0$ (i.e. $p=0$) is a proper
eigenvalue (although $p=0$ is not a zero of (9)) to which the trivial null scattering eigenfunction corresponds.
For IBCs (or OBCs) the eigenfunctions corresponding to $p_+$  (or $p_-$) exist but are null.

As noticed above, there exists a crisscrossed genetic kinship between the
\emph{in} (\emph{out}) scattering solutions and the outgoing
(incoming) resonances defined by the OBCs (IBCs), including the
anti-bound (bound) states, rendering the extension of the former to the
complex plane a little messy \cite{Madrid2}. For complex momenta the terms \emph{incoming}
and \emph{outgoing} loose much physical meaning, which maybe related,
at most, to the sign of the real part of the momenta. This relationship
breaks down for the resonant (either outgoing or incoming) solutions,
which have momenta with both positive and negative real parts, and it is fully
meaningless for the bound and anti-bound solutions, which have zero real part.
The OBCs (IBCs) rather determine the location of the zeroes in ${\cal C}_-$
(${\cal C}_+$) . In fact, it can be shown (Appendix A)
that the OBCs (IBCs) imply that the momenta of the solutions feature
a negative (positive) imaginary part, regardless the real part being
zero (bound and anti-bound states) or finite (resonances).

The zeroes of the denominators (9) can be pinpointed only by
numerically computing the roots of ${\cal D}_+(p)=0$ or ${\cal
D}_-(p)=0$ . However the naive use of mathematical programs may
yield spooky results that can be traced back to the default used for
the branching of the Square Root function, so that $\sqrt{z^2}$ not
always yields $z$ . For instance, in the case of ${\cal D}_+=0$ for the
square well in (9), one may find the expected resonance zeroes in ${\cal C}_-$
 together with only \emph{part} of the anti-bound momenta in
the lower imaginary axis \emph{plus} some bound state momenta (upper
axis). The remaining ones, together with the incoming resonant
momenta in ${\cal C}_+$ , stem from ${\cal D}_-=0$, the roots
of which are the complex conjugates of ${\cal D}_+=0$.

To conclude this section, an overview of the location in the complex
plane of the momenta corresponding to the different types of solutions
above is given in Figure 1.

\section{Inner products}

From the MS (8) we may obtain a Master Inner Product (MIP)
\begin{equation}
\langle{\cal E}|{\cal E'}\rangle=\int^{+\infty}_{-\infty}\rmd
x\,\Phi^*_{\cal E}(x)\Phi_{\cal E'}(x)
\end{equation}
where the labels $+,r$ are understood and will be omitted in the following.
The integral can be computed along the usual lines (Appendix B), yielding
\begin{equation}
\langle {\cal E}|{\cal
E'}\rangle={\cal N}^*{\cal N}'\left\{
\begin{array} {l}
\;\;\;[I(z^*-z')-\frac{\rmi}{z^*-z'}]\,{\cal D}^*{\cal D}'\\
+[I(-(z^*-z'))+\frac{\rmi}{z^*-z'}]\,({\cal R}^*{\cal R}'+{\cal T}^*{\cal T}')\\
+[I(-(z^*+z'))+\frac{\rmi}{z^*+z'}]\,{\cal R}^*{\cal D}'\\
+[I(z^*+z')-\frac{\rmi}{z^*+z'}]\,{\cal D}^*{\cal R}'\\
\end{array}
\right.
\end{equation}
where $I(k)\equiv\int_{0}^{\infty}\rmd x\,e^{\rmi kx}$, and the primed notation is explained in that Appendix. All the inner products of the different families of solutions
with themselves, and with each other, can be derived from (11), and in the remaining of this section
we shall outline the technicalities of the most relevant cases.

\subsection{Scattering states}
For real $z=p$ and $z'=p'$, we readily see that the MIP (11) yields $|{\cal
D}|^2\langle E|E'\rangle$ :
\begin{equation}
\fl\langle {\cal E}|{\cal
E'}\rangle={\cal D}^*{\cal D}'\langle E|E'\rangle={\cal D}^*{\cal D}'
N^*N'\left\{
\begin{array} {l}
\;\;\;[I(p-p')-\frac{\rmi}{p-p'}]\\
+[I(-(p-p'))+\frac{\rmi}{p-p'}]\,(R^*R'+T^*T')\\
+[I(-(p+p'))+\frac{\rmi}{p+p'}]\,R^*\\
+[I(p+p')-\frac{\rmi}{p+p'}]\,R'\\
\end{array}
\right.
\end{equation}
For real $k$, the integrals $I(k)$ are singular (Appendix C) and must be interpreted as a distribution,
namely
\begin{equation}
I(\pm k)=\pm\rmi\,PV\frac{1}{k}+\pi
\,\delta(k)
\end{equation}
As a result, all the terms proportional to $(p\pm p')^{-1}$ stemming from these Principal Values,
plus the ones displayed in (12), cancel out. The terms $\delta(p+p')$ vanish because $p+p'$ is always $>0$, and the terms $\delta(p-p')$ sum up to
\begin{equation}\fl
\langle E|E'\rangle=\frac{1}{2\pi}\frac{m}{\sqrt{p\,p'}}\;\pi\,\delta(p-p')\,[1+|R|^2+|T|^2]
=\frac{m}{p}\,\delta(p-p')=\delta(E-E')
\end{equation}
as expected from the wave function normalization adopted.

\subsection{Background states}

For general complex $z$ and $z'$, the general form of $I(k)$ (Appendix C) leads to the result
\begin{equation}
\langle {\cal E}|{\cal E'}\rangle= \left\{
\begin{array}{ll}
 0&\quad,\quad-\frac{\pi}{4}< {\rm arg}(z'-z^*)<5\frac{\pi}{4}\\
\infty&\quad,\quad {\rm otherwise}
\end{array}\right.
\end{equation}
Thus, for each background state $|\,{\cal E}\rangle$ of momentum $z$, there is
a "neighborhood of divergence" so that $|\,{\cal E}\rangle$ is orthogonal to
any other $|\,{\cal E'}\rangle$ the momentum $z'$ of which lies outside a
"divergence wedge", with an angle of $\pi/4$ and apex in $z^*$, and gives a divergent
inner product if $z'$ lies inside this wedge. The rule is reciprocal:
we may consider as well the location of $z$ with respect to the wedge with apex
in $z'^*$. In particular $\langle {\cal E}|\,{\cal E}\rangle= \infty$, as expected.
When both $z$ and $z'$ lie on the real axis, both apexes fall on the real axis
so that the only divergent products happens for $z=z'$, as already known for the
scattering states.

The products with other states yield highly variable results
depending on the location of $z$ on the whole complex plane. Of most
interest is the sector $7\frac{\pi}{4}<{\rm arg}(z)<2\pi$, in which
case the states $|\,{\cal E}\rangle$ are orthogonal to the bound states and
partially orthogonal to the anti-bound and to the scattering states.

\subsection{Bound and Anti-bound states}

As explained in $2.4.$ , the MS (8) stems from the analytical continuation of the scattering
$in$ solutions we are mainly considering. Then the zeroes of ${\cal D}_+(z)$ lie only in ${\cal C}_-$
and do not include the bound state momenta. However, apart from the specific form of the amplitudes
${\cal R}$ and ${\cal T}$, both bound (which would stem from the scattering $out$ solutions, then corresponding to zeroes of ${\cal D}_-(z)$, lying in ${\cal C}_+$) and
anti-bound solutions share the general form (8) with respectively ${\cal D}_\mp=0$.
For ease of writing, in this subsection we drop the tilde notation introduced in 2.1.
Thus we consider bound states $|\phi_i\rangle$ with purely imaginary momenta $z_i={\rm i}q_i$
($q_i>0$), while the anti-bound states have $q_i<0$.
The general inner products $\langle \phi_i|\phi_j\rangle$ stem from (11) with the simplification
${\cal D}={\cal D}'=0$.

The bound states involve only finite integrals.
For $i\neq j$ we have
\begin{equation}
\langle \phi_i|\phi_j\rangle=\frac{1}{2\pi}\frac{m}{\sqrt{q_iq_j}}\;({\cal R}^*_i{\cal R}_j
+{\cal T}^*_i{\cal T}_j)\;[\,I(\rmi(q_i+q_j))-\frac{1}{q_i+q_j}\,]=0
\end{equation}
since
\begin{equation}
I(\rmi(q_i+q_j))=\int_{0}^{\infty}\rmd x\,e^{-(q_i+q_j)x}=\frac{1}{q_i+q_j}
\end{equation}
It is instructive to notice that (B.3) becomes
\begin{eqnarray}
\fl-(q^2_i-q^2_j)\int^L_0\rmd
x\,\phi^*_i\phi_j&=&2m(E_i-E_j)\int^L_0\rmd x\,\phi^*_i\phi_j\\\nonumber
&=&W[\phi^*_i, \phi_j]^L_0=(q_i-q_j)\frac{1}{2\pi}\frac{m}{\sqrt{q_iq_j}}\,({\cal R}^*_i {\cal R}_j+{\cal T}^*_i {\cal T}_j\,e^{-(q_i+q_j)L})\;,
\end{eqnarray}
and (B.4) now is
\begin{equation}
\int^L_0\rmd
x\,\phi^*_i\phi_j=-\frac{1}{(q_i+q_j)}\frac{1}{2\pi}\frac{m}{\sqrt{q_iq_j}}\,({\cal R}^*_i{\cal R}_j+{\cal T}^*_i{\cal T}_j\,e^{-(q_i+q_j)L})\;.
\end{equation}

For $i=j$ , (18) is useless for the calculation of $\int^L_0\rmd
x\,\phi^*_i\phi_i$ . Instead the (finite) norm
$\|\phi_i\|^2=\langle\phi_i|\phi_i\rangle$ can be obtained by using
the Wronskian $W[\phi_i^*,\partial_{q_i}\phi_i]$ (see
\cite{Berggren3} for this "method of quadratures", say integration
by parts, applied to resonances). Noticing that
$\partial^2_x\,\phi_i=[2mV(x)+q^2_i]\,\phi_i$ and that
$\partial^2_x\,\,\partial_{q_i}\phi_i=[2mV(x)+q^2_i]\,\partial_{q_i}\phi_i+2q_i\phi_i$
, we now have
\begin{equation}
2q_i\int^L_0\rmd x\,\phi^*_i\phi_i=W[\phi^*_i,
\partial_{q_i}\phi_i]^L_0\;.
\end{equation}
On the other hand, using the BCs at $x=0,L$ , the terms involving
the derivatives $\partial_{q_i}{\cal R}_i$ and $\partial_{q_i}{\cal T}_i$ cancel
out so that we now have
\begin{equation}
W[\phi^*_i,
\partial_{q_i}\phi_i]^L_0=\frac{1}{2\pi}\frac{m}{q_i}\,(|{\cal R}_i|^2+|{\cal T}_i|^2\rme^{-2q_iL})\;.
\end{equation}
and hence
\begin{equation}
\int^L_0\rmd x\,\phi^*_i\phi_i=\frac{1}{2\pi}\frac{m}{2q_i^2}\,(|{\cal R}_i|^2+|{\cal T}_i|^2\rme^{-
2q_iL})\;.
\end{equation}
(notice that the inadvertent use of (19) with $i=j$ yields (22) albeit for
a crucial global sign). Then
\begin{eqnarray}\nonumber
&\fl\quad\quad\langle\phi_i|\phi_i\rangle&=\frac{1}{2\pi}\frac{m}{q_i}\,((|{\cal R}_i|^2+|{\cal T}_i|^2)\int_0^{\infty}\rmd
x\,e^{-2q_ix}-|{\cal T}_i|^2\int_0^L\rmd x\,e^{-2q_ix})+\int_0^L\rmd
x\,\phi^*_i\phi_i\\\nonumber
&\fl\quad&=\frac{1}{2\pi}\frac{m}{2q_i^2}\,(|{\cal R}_i|^2+|{\cal T}_i|^2e^{-2q_iL})+\int_0^L\rmd
x\,\phi^*_i\phi_i\\
&\fl\quad&=\frac{1}{2\pi}\frac{m}{q_i^2}\,(|{\cal R}_i|^2+|{\cal T}_i|^2e^{-2q_iL})\;,
\end{eqnarray}
which yields the wave function normalization factor needed for having
$\langle \phi_i|\phi_j\rangle=\delta_{ij}$ .

The inner products of the anti-bound states between themselves are divergent, as expected.
However, their products with the bound states are dominated in some cases by the convergent
spatial behaviour of the latter, in which case we have orthogonality. From (17) we see that
this happens whenever $(q_i+q_j)>0$, the integral being divergent otherwise.

Likewise, the convergent power of the bound states may dominate over the divergent behaviour of
the general background solutions, so that a partial "wedge" orthogonality governs also these crossed
inner products.

The text book orthogonality between the bound and the scattering states can be immediately read out from (11).
For the product $\langle \phi_i|E\rangle$ one has ${\cal D}=0$, the integrals $I(\rmi q_i\pm p)$ are finite and yield
$\rmi(\rmi q_i\pm p)^{-1}$ respectively, so that we have $\langle \phi_i|E\rangle=0$.

\subsection{Resonant and Scattering states}

Consider a resonance with momentum $p_n$ and a scattering $in$ state with real momentum $p>0$. Omitting an irrelevant normalization factor ${\cal N}^*_nN$ , then (11) reduces to
\begin{equation}\fl\langle
z_n|E\rangle\propto[I(-(p^*_n-p))+\frac{\rmi}{p^*_n-p}]\,({\cal R}^*_n{\cal R}+{\cal T}^*_n{\cal T})+[I(-(p^*_n+p))+\frac{\rmi}{p^*_n+p}]\,{\cal R}^*_n{\cal D}(p)
\end{equation}
With $p>0$ , for $n>0$ we always have $-\frac{\pi}{4}< {\rm
arg}(-(p^*_n+p))<5\frac{\pi}{4}$ , so that
\begin{equation}\langle
z_n|E\rangle\propto[I(-(p^*_n-p))+\frac{\rmi}{p^*_n-p}]\,({\cal R}^*_n{\cal R}+{\cal T}^*_n{\cal T})
\end{equation}
and therefore, with the prescription adopted,
\begin{equation}
\langle z_n|E\rangle= \left\{
\begin{array}{ll}
 0&\quad,\quad-\frac{\pi}{4}< {\rm arg}(p-p^*_n)<5\frac{\pi}{4}\\
\infty & \quad,\quad{\rm otherwise}
\end{array}\right.
\end{equation}
This means that a given scattering \emph{in} state $|E\rangle$ (with
momentum $p>0$ on the real axis) is orthogonal to any $|z_n\rangle$
(with $n>0$) if the momentum $p_n$ lies outside the wedge with apex
in $p$ , the inner product being divergent otherwise. Viceversa,
given $p_n$ , the momenta $p\,$ of the orthogonal scattering states
lie outside the wedge with apex in $p_n$ .

The scattering states have a reflected wave with momentum of
opposite sign to the incident one, so the situation is trickier for
$n<0$ . Given $p_{-|n|}$ , the wedge to be considered is again the
one with apex in the mirror momentum $p_{\,|n|}$ .

\subsection{Resonant and bound states}

Up to a normalization factor, the MIP now reduces to
\begin{equation}\langle\phi_i|z_n\rangle\propto({\cal R}^*_i{\cal R}_n+{\cal T}^*_i{\cal T}_n)[I(\rmi q_i+p_n)-\frac{\rmi}{\rmi q_i+p_n}]
\end{equation}
The integral is convergent, yielding $\rmi(\rmi q_i+p_n)^{-1}$ and
hence $\langle\phi_i|z_n\rangle=0$, provided that $q_i>|{\rm
Im}\,p_n|$. We thus have a situation similar to that occurring
between resonances, namely that the states are orthogonal if the
bound state momentum, lying in the positive imaginary axis, and the
resonant one, lie outside the respective divergence wedges, the
inner product being infinite otherwise. For a potential well of
finite depth featuring both bound and resonant states, the latter
case happens only for a few of the (finite number of) bound states.

\subsection{Resonances and anti-bound states}

As long as we start from the solution (8), which satisfies an OBC, and from the
corresponding MIP (11), in this subsection we consider the inner products of the
states with isolated momenta lying in the lower half complex plane, namely the
outgoing resonances with themselves and with the anti-bound solutions.
The result is similar to (15), albeit for the fact that the momenta $z$ and $z'$
take the isolated denumerable values $\rmi q_i$ or $z_n$ :

\begin{equation}
\langle {a}|{b}\rangle= \left\{
\begin{array}{ll}
 0&\quad,\quad-\frac{\pi}{4}< {\rm arg}(z_b-z^*_a)<5\frac{\pi}{4}\\
\infty&\quad,\quad {\rm otherwise}
\end{array}\right.
\end{equation}
where the labels $a$ and $b$ run over the the values of $i$ or $n$.
A similar layout of "divergence wedges" comes out.

Through this section the structure of divergence wedges above describes
a partial orthogonality between the members of the different families of
solutions of the SE. We shall call it "wedge orthogonality" and indicate
it by the symbol $\Delta_{\,ab}$ . In the Figure 1 we have depicted the
divergence wedge of the resonant state $|z_2\rangle$. Examples
involving resonant and scattering solutions can be found in Figure 2 in \cite{Julve1}.

\begin{figure}[h]
\begin{center}
\includegraphics[width=0.8\textwidth]{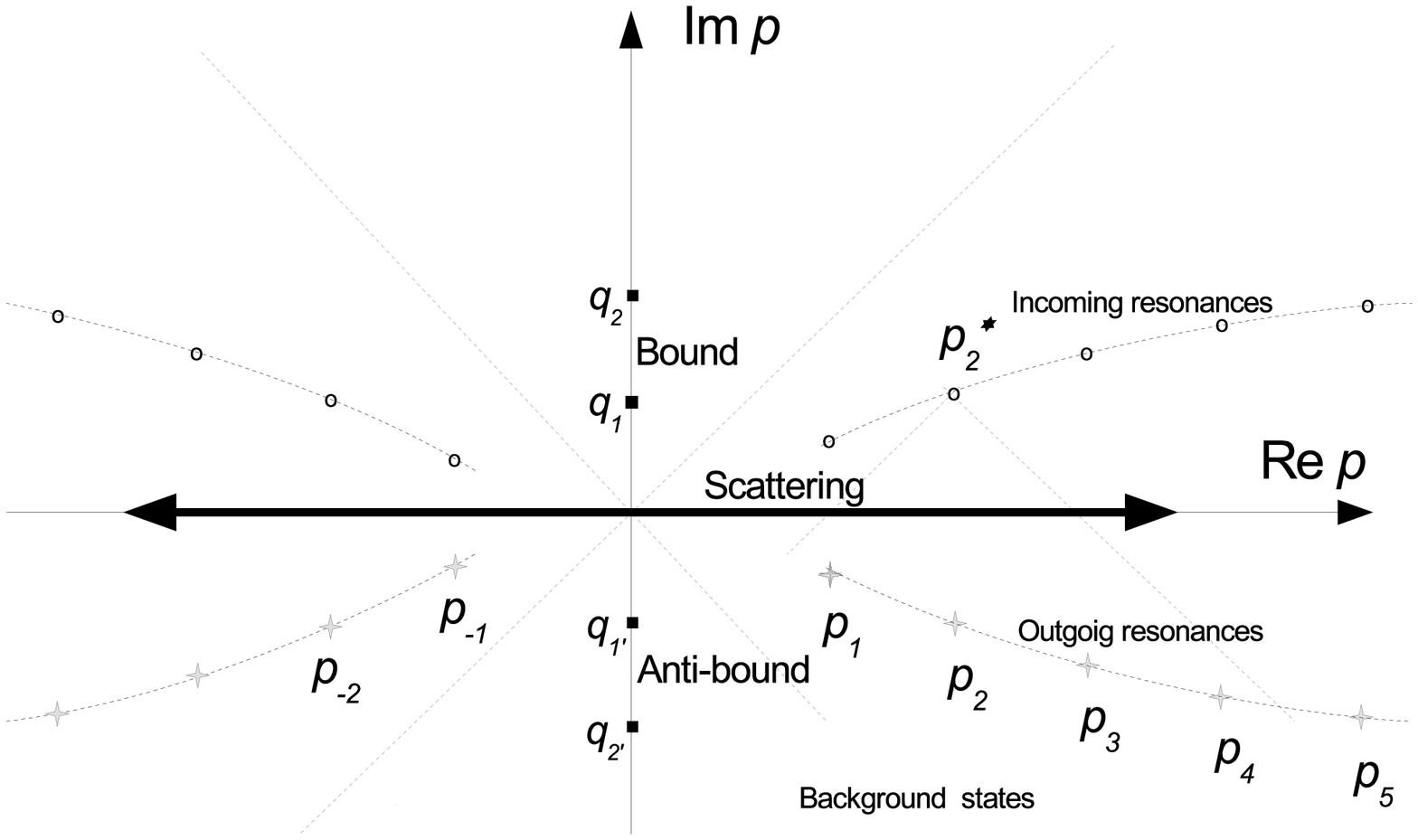}
\caption{Location of the momenta of the solutions.\hfill\break
Two bound states are assumed. An example of "wedge orthogonality" \hfill\break
is given between resonant
states: $\langle z_2|z_n\rangle=0$ for $n<0$ and $n>4$ , \hfill\break
while $\langle z_2|z_m\rangle=\infty$ for $0<m\leq4$ .}
\label{Fig.1}
\end{center}
\end{figure}

\subsection{General products}

All the inner products considered above may be displayed in the
following comprehensive table:
\begin{center}
  \begin{tabular}{r||c|c|c|c|c|c|c|c|c|}
    $\langle a|\,b\,\rangle$ & $|\phi_2\rangle$
      & $|\phi_1\rangle$ & $|\phi_{1'}\rangle$
      & $|\phi_{2'}\rangle$ & $|z_m\rangle$ & $|\tilde z_m\rangle$ & $|E'\rangle$ & $|{\cal E}'\rangle$ \\
      \hline \hline
    $\langle\phi_2|$ & $1$ & $0$ & $0$ & $\infty$ & $0$ & $0$ & $0$ & $\Delta_{\,ab}$ \\ \hline
    $\langle\phi_1|$ & $0$ & $1$ & $\infty$ & $\infty$ & $0$ & $0$ & $0$ & $\Delta_{\,ab}$ \\ \hline
    $\langle\phi_{1'}|$ & $0$ & $\infty$ & $\infty$ & $\infty$ & $\Delta_{\,ab}$ & $\Delta_{\,ab}$ & $\Delta_{\,ab}$ & $\Delta_{\,ab}$ \\ \hline
    $\langle\phi_{2'}|$ & $\infty$ & $\infty$ & $\infty$ & $\infty$ & $\Delta_{\,ab}$ & $\Delta_{\,ab}$ & $\Delta_{\,ab}$ & $\Delta_{\,ab}$ \\ \hline
    $\langle z_n|$ & $0$ & $0$ & $\Delta_{\,ab}$ & $\Delta_{\,ab}$ & $\Delta_{\,ab}$ & $0$ & $\Delta_{\,ab}$ & $\Delta_{\,ab}$ \\ \hline
    $\langle\tilde z_n|$ & $0$ & $0$ & $\Delta_{\,ab}$ & $\Delta_{\,ab}$ & $0$ & $\Delta_{\,ab}$ & $0$ & $\Delta_{\,ab}$ \\ \hline
    $\langle E|$ & $0$ & $0$ & $\Delta_{\,ab}$ & $\Delta_{\,ab}$ & $\Delta_{\,ab}$ & $0$ & $\delta(E-E')$ & $\Delta_{\,ab}$ \\ \hline
    $\langle {\cal E}|$ & $\Delta_{\,ab}$ & $\Delta_{\,ab}$ & $\Delta_{\,ab}$ & $\Delta_{\,ab}$ & $\Delta_{\,ab}$ & $\Delta_{\,ab}$ & $\Delta_{\,ab}$ & $\Delta_{\,ab}$ \\ \hline
  \end{tabular}
\end{center}

\noindent{Here} a barrier including a well of finite depth supporting two
ordinary bound states $|\phi_1\rangle$ and $|\phi_2\rangle$ with
energies $|E_1|<|E_2|$ , plus the mirror anti-bound states
$|\phi_{1'}\rangle$ and $|\phi_{2'}\rangle$, has been assumed, with
the ordering $q_2>q_1>0>q_{1'}>q_{2'}$ . The value of the products
involving the background states $|{\cal E}\rangle$ depends on their
momentum $z$, the range of which is the whole ${\cal C}$.
Here we have considered $z$ and $z'$ in ${\cal C}_-$.

\section{Conclusions}

We have calculated the relevant inner products involving the
resonant eigenstates and the more general complex energy "background"
solutions in the example of a 1-dimensional quantum
potential barrier with compact support. As it is well known, these
states respectively have discrete and continuum complex energies and,
in the space spanned by them, the Hamiltonian is not Hermitean, so that
the usual neat results about the (either Dirac or Kronecker-$\delta$)
orthogonality, the reality of the norms, etc., do not hold. On the other
hand, their modulus grows exponentially at the spatial infinity, giving
rise to infinite norms and seemingly infinite inner products.

Among the variety of proposals to circumvent these difficulties, we
have adopted a Gaussian convergence factor, first introduced by Zel'dovich,
and carried out the limit of the integrals where the factor fades
off to unity. This prescription yields inner products such that most
of these states are orthogonal to each other, except when they lie
in a neighborhood described by a "divergence wedge" in the complex-momentum
plane, in which case the product is infinite.

The guiding thread of this paper relies on the analytic continuation
to the momentum complex plane ${\cal C}$ of the scattering solutions.
In doing so, we have started from an $in$, right-moving, solution
(obeying an Outgoing Scattering Boundary Condition), from which a Master
Solution is obtained. This MS describes the whole family of solutions
when the momentum takes particular values in ${\cal C}$, namely the
($in$, $r$) scattering states for $z=p\in{\cal R}_+$, the anti-bound states
for isolated $z_i\in{\cal I}_-$, the outgoing resonances (obeying two Outgoing
BCs) for isolated $z_n\in{\cal C}_-$, and the continuum of the general
background solutions which are holomorphic in the remaining of ${\cal C}$.
An alternative MS can be defined starting from the $out$ scattering states
(obeying an Incoming Scattering BC), which gives rise to the bound states and to
the incoming resonances (obeying two Incoming BCs), all of them with momenta
$\in{\cal C}_+$.

An important result is that the outgoing (incoming) BCs dictate that the solution's
momenta $\in{\cal C}_-$ ($\in{\cal C}_+$) in full generality, that is
regardless their real part being finite (resonances) or null (anti-bound or
bound states). This makes the outgoing resonances akin to the anti-bound states
(and the incoming ones to the bound states), whereas there is a somehow messy
crossed kinship between the outgoing (incoming) BCs and the parent $in$ ($out$) MS.

Out of the MS above a Master Inner Product has been defined, from which all the
inner products between the members of the different families of solutions stem.
Generally a partial "wedge orthogonality" is obtained, which we denote by the symbol
$\Delta_{\,ab}$ , from which the expected infinite norms of all the solutions, except
for the bound states, results.
Of course, the traditional orthogonality relations between the (real energy)
bound and scattering states, namely the Kronecker-$\delta$ and the Dirac-$\delta$,
arise as particular cases.
This results is different from the full bi-orthogonality obtained by the prescription
of analytically continuing the finite integrals from ${\cal C}_+$ to the whole
${\cal C}_-$ , where the integrals are formally divergent. Our limiting procedure
instead extends the finite result to only the $\pi/4$ wedges of ${\cal C}_-$ close
to the real axis.

\ack{Work supported by MEC project FIS2011-29287. J. Julve acknowledges the hospitality
of the Dipartimento di Fisica dell'Universit\`a di Bologna, Italy, where part of this
work was done.}

\appendix

\section{Boundary conditions and pole location}

We work out the relationship between the OBCs, the IBCs, and the location
of the momenta in the complex plane.  We first re-derive for 1-D quantum systems
the known link between the OBCs and the resonances in the lower half plane
\cite{Nussenzveig} \cite{Calderon2}.

We re-write (1) in the more convenient form
\begin{equation}
\partial_x^2u(x)= [2mV(x)-p^2]\,u(x)
\end{equation}
and consider the OBCs (4) together with their complex conjugates.
Computing the expression
\begin{equation}
\int^L_0\rmd x\;[u^*\partial_x^2u-u\,\partial_x^2u^*]
\end{equation}
both using (A.1) and integration by parts, yields the equality
\begin{equation}
(p^{*2}-p^2)\int^L_0\rmd x\;|u|^2=\rmi(p+p^*)(|u(0)|^2+|u(L)|^2)
\end{equation}
With $p=\alpha+\rmi\beta$ one obtains
\begin{equation}
-\rmi 4\alpha\beta\int^L_0\rmd x\;|u|^2=\rmi 2\alpha(|u(0)|^2+|u(L)|^2)\;,
\end{equation}
which, for $\alpha\neq 0$ , yields
\begin{equation}
\beta=-\frac{|u(0)|^2+|u(L)|^2}{2\int^L_0\rmd x\;|u|^2}\;\;<\,0\;.
\end{equation}

The case $\alpha=0$ (anti-bound solutions) may be dealt with by using the momentum-derivative
$u'(x)\equiv\partial_pu(x)$ , which obeys the equations
$\partial_x^2u'= [2mV-p^2]\,u'-2p\,u$ and $\partial_xu'|_{\,0}=-\rmi p\,u'(0)-\rmi\,u(0)\;,\;\partial_xu'|_L=\rmi p\,u'(L)+\rmi\,u(L)\;,$
and starting from the expression
\begin{equation}
\int^L_0\rmd x\;[u^*\partial_x^2u'-u'\,\partial_x^2u^*]\;.
\end{equation}
Then we obtain the equality
\begin{eqnarray}
\fl-\,\rmi\,4\alpha\beta\int^L_0\rmd x\;u^*u'-&2(\alpha+\rmi\beta)\int^L_0\rmd x\;|u|^2\\\nonumber&=\;\rmi\,2\alpha
(u^*(0)u'(0)+u^*(L)u'(L))+\rmi\,(|u(0)|^2+|u(L)|^2)\nonumber
\end{eqnarray}
which, for $\alpha=0$ , yields the same result (A.5).

The IBCs lead to $\beta>0$ for any $\alpha$, and give rise to both the incoming resonances and the bound states.

\section{Master Inner Product}

The space integral in (10) can be split in three sectors
\begin{equation}\fl
\langle{\cal E}|{\cal E'}\rangle={\cal N}^*{\cal N}'\left\{
\begin{array} {l}
\int^{\;0}_{-\infty}\rmd
x\,[{\cal D}^*{\cal D'}e^{-\rmi(z^*-z')x}+{\cal R}^*{\cal R'}e^{\rmi(z^*-z')x}\\
\quad\quad\quad+{\cal R}^*{\cal D'}e^{\rmi(z^*+z')x}+{\cal D}^*{\cal R'}e^{-\rmi(z^*+z')x}]\\
+({\cal N}^*{\cal N}')^{-1}\int^L_0\rmd
x\,\Phi^*_{\cal E}(x)\Phi_{\cal E'}(x)\\
+\int^{\infty}_L\rmd
x\,{\cal T}^*{\cal T'}e^{-\rmi(z^*-z')x}\\
\end{array}
\right.
\end{equation}
with the short-hand notation ${\cal N}^*\equiv({\cal N}(z))^*$ , ${\cal N'}\equiv{\cal N}(z')$ , and similarly for ${\cal D}$ and the amplitudes ${\cal R}$ and ${\cal T}$.
We then bring all the infinite-limited integrals to the form $I(k)=\int_{0}^{\infty}\rmd
x\,e^{\rmi kx}$ , so that
\begin{equation}\fl
\langle {\cal E}|{\cal E'}\rangle={\cal N}^*{\cal N}'\left\{
\begin{array} {l}
I(z^*-z'){\cal D}^*{\cal D'}+I(-z^*+z')({\cal R}^*{\cal R'}+{\cal T}^*{\cal T'})\\
+I(-(z^*+z')){\cal R}^*{\cal D'}+I(z^*+z'){\cal D}^*{\cal R'}\\
+({\cal N}^*{\cal N}')^{-1}\int^L_0\rmd x\,\Phi^*_{\cal E}(x)\Phi_{\cal E'}(x)\\
-{\cal T}^*{\cal T'}\int^L_0\rmd x\,e^{-\rmi(z^*-z')x}\\
\end{array}
\right.
\end{equation}
The form (8) of the solution at $x=0$ and at $x=L$ lets expressing $\int^L_0\rmd x\,\Phi^*_{\cal E}\Phi_{\cal E'}$ in
terms of the amplitudes ${\cal D}$, ${\cal R}$ and ${\cal T}$ outside the barrier. The procedure uses the
operator $\partial^2_x$ for which $\partial^2_x\Phi_{\cal E}=2m(V-{\cal E})\Phi_{\cal E}$ , and
integration by parts to obtain
\begin{eqnarray}
\fl(z^{*2}-z'^2)\int^L_0\rmd
x\,\Phi^*_{\cal E}\Phi_{\cal E'}&=&2m({\cal E}^*-{\cal E'})\int^L_0\rmd x\,\Phi^*_{\cal E}\Phi_{\cal E'} \\\nonumber
&=&\int^L_0\rmd
x\,\Phi_{\cal E}^*(x)(\overrightarrow{\partial^2_x}-\overleftarrow{\partial^2_x})\Phi_{\cal E'}(x)=W[\Phi^*_{\cal E},
\Phi_{\cal E'}]^L_0\\\nonumber
&=& \rmi(z^*+z'){\cal N}^*{\cal N}'(-{\cal D}^*{\cal D'}+{\cal R}^*{\cal R'}+{\cal T}^*{\cal T}'e^{-\rmi(z^*-z')L})\\\nonumber
&&\!\!+\rmi(z^*-z'){\cal N}^*{\cal N}'({\cal R}^*{\cal D'}-{\cal D}^*{\cal R}')\;\;,
\end{eqnarray}
where $W[\phi\,,\psi]\equiv
\phi\,\partial_x\psi-\psi\partial_x\,\phi$ is the Wronskian, so that
\begin{equation}\fl\int^L_0\rmd
x\,\Phi^*_E\Phi_{E'}=\frac{\rmi\,{\cal N}^*{\cal N}'}{z^*-z'}(-{\cal D}^*{\cal D'}+{\cal R}^*{\cal R'}+{\cal T}^*{\cal T}'e^{-\rmi(z^*-z')L})+\frac{\rmi\,{\cal N}^*{\cal N}'}{z^*+z'}({\cal R}^*{\cal D'}-{\cal D}^*{\cal R}')
\end{equation}
provided that ${\cal E}^*\neq{\cal E}'$ (i.e. $z^{*2}\neq z'^2$).
The finite last integral term in (B.2) gives the result $\rmi
(z^*-z')^{-1}{\cal T}^*{\cal T}'(1-e^{-\rmi(z^*-z')L})$ , which readily leads to (11).

The case ${\cal E}^*={\cal E}'$ arises in the calculation of the norms whether they be finite (bound states) or divergent (all the others). In the latter case, the infinite result shows up already in the integrals $I(k)$ in (B.2), regardless of the method used to compute the finite integral $\int^L_0\rmd x\,\Phi^*_{\cal E}(x)\Phi_{\cal E'}(x)$. We calculate the finite result of the bound states in 3.3 by using the technique of quadratures.

\section{Infinite-limited integrals}

For computing $I(k)$ we rely on the limit $\lambda\rightarrow 0$ of the basic Gaussian regularized integral
\begin{equation}
J(k,\lambda)\equiv\int^{+\infty}_0 \rmd x\; \rme^{-\lambda
x^2}\rme^{\rmi kx }=\frac{\rmi}{k}\;\sqrt{\pi}\,\tau\,\rme^{\tau^2} {\rm
erfc}(\tau)\hskip 1.0cm (\lambda\;{\rm real}>0)
\end{equation}
which is directly related to (7.1.2) in \cite{Stegun}, where
$\tau=-\rmi k/(2\sqrt{\lambda})$ , and hence $k$ , can take any complex
value. See Appendix A in \cite{Julve1} for more details.

For Im $k\;>0$ the integral is always convergent so that the limit can be taken
in the integrand in (C.1). Then we trivially have $I(k)\equiv J(k,0)=\rmi\,k^{-1}$.

For real $k$ , the quoted result
\begin{equation}
\int^{\infty}_0\rmd x \, e^{\rmi kx}= \rmi\,PV\frac{1}{k}+\pi
\,\delta(k)
\end{equation}
relies on adding to $k$ a small imaginary part $\rmi\epsilon$, which
still guarantees the convergence when $\lambda\rightarrow 0$, but later
in the limit $\epsilon\rightarrow 0_+$ the result must be interpreted
as a distribution.

For Im $k\;<0$ the integration and the limit $\lambda\rightarrow 0$
do not commute and we adopt the limit of the integral as a
prescription. From (7.1.23) in \cite{Stegun} we obtain
\begin{equation}
 I(k)\equiv J(k,0)=\left\{
\begin{array}{ll}
\frac{\rmi}{k}&\quad ,\quad
-\frac{\pi}{4}<{\rm arg}(k)<5\frac{\pi}{4} \;\;\;,\;\; k\neq 0\\
\infty & \quad , \quad{\rm otherwise}
\end{array}
\right.
\end{equation}
which extends the finite result $\rmi k^{-1}$ to a new region of ${\cal C}_-$ and defines the "divergence wedge" there mentioned.

\end{document}